\documentstyle[12pt]{article}
\begin{document}
\title{Mesoscopic Mechanics\footnote{This work has been presented in part at the
International Symposium on Inhomogeneous and Strongly Correlated Materials
with Novel Electronic Properties (ISCM), held at Miami Beach FL (USA) as a
part of the SMEC meeting, during March 24-28, 2003.}}
\author{Artur Sowa\\
109 Snowcrest Trail,
Durham, NC 27707 \\
www.mesoscopia.com }
\date{}
\maketitle
\begin{abstract}

This article is concerned with the existence, status and description of the so-called emergent phenomena believed to occur in certain principally planar electronic systems. In fact, two distinctly different if inseparable tasks are accomplished. First, a rigorous mathematical model is proposed of emergent character, which is conceptually bonded with Quantum Mechanics while apparently non-derivable from the many-body Schr\"{o}dinger equation. I call the resulting conceptual framework the Mesoscopic Mechanics (MeM). Its formulation is space-independent and comprises a nonlinear and holistic extension of the free electron model. Secondly, the question of relevancy of the proposed ``emergent mechanics" to the actually observed phenomena is discussed. In particular, I postulate a probabilistic interpretation, and indicate how the theory could be applied and verified by experiment. 

The Mesoscopic Mechanics proposed here has been deduced from the Nonlinear Maxwell Theory (NMT)---a classical in character nonlinear field theory. This latter theory has already been shown to provide a consistent phenomenological model of such phenomena as superconductivity, charge stripes, magnetic vortex lattice, and magnetic oscillations. The NMT, which arose from geometric considerations, has long been awaiting an explanation as to its ties with the fundamental principles. I believe the MeM provides at least a partial explanation to this effect.

\end{abstract}

\section{The incentive to consider emergent mathematical structures}
\label{emerge}

The article \cite{Laughlin-Pines} by R. B. Laughlin and David Pines inspires us to re-examine the status of the reductionist paradigm, and perhaps also the basic assumptions as to the meaning of {\em complexity} and its bearing on how we attempt to understand physical phenomena. Moreover, an even more bewildering question stands whether complexity may play a direct role in how things actually work. Still more disturbingly, the only meaning of the phrase ``how things work" may be endowed to it by our attempt to understand how they work, and depend on its peculiarities including our choice of the paradigm. In particular, there may be several mathematically nonequivalent ways of modelling one and the same phenomenon, all true, coexisting alongside and complementing one another. Be it as it may, there is a danger of dogmatism in rejecting any possibility that there is room for {\em emergent} way of thinking and {\em emergent} mathematical models.   

The beautiful and puzzling phenomena observed in some new principally planar electronic systems based on novel materials invite reflection on justifiability of a functionally apt mathematical model that does not exactly begin with the Schr\"{o}dinger equation. Before anything else, it may be worthwhile to try and examine the ``mathematical reality" of this problem and ask if we can construct mathematical models of clearly  {\em emergent} character at all. A model of this sort would have to be conceptually tied with the basic principles while impossible or prohibitively difficult to deduce from the Schr\"{o}dinger equation in conjunction with the Pauli exclusion principal. These singular requirements seem quite reasonable inasmuch as they parallel our intuitive grasp of the nature of the phenomena suspected of being {\em emergent}. It is impossible not to mention some preexisting examples that at least come close to fulfilling these requirements, like the classical Ginzburg-Landau equation or the Solitons. Both these theories may be viewed as supplying {\em emergent} models, save their ties with the basic principles are in a typical application postulated rather than inherent or rigorously derived. In this sense, these classical theories are not perfect examples of emergent mathematical models.

Many macroscopic electronic properties of 2D electron gas in a magnetic field depend on the following basic if somewhat idealized and simplified picture resulting from the free electron model, which is good to keep in mind during our discussion. Namely, as we apply perpendicular to the sample increasing magnetic field $B$, the separation between Landau levels increases proportionately. As it happens, the consecutive Landau levels cross over the Fermi level and some electrons residing at these levels are emptied while some reoccupy a lower Landau level. This contributes directly to the conductivity of the sample. In particular, the longitudinal conductivity, as well as other macroscopic parameters, will display oscillatory dependence on the magnetic induction. At low temperature, the number of electrons occupying each Landau level is close to the degeneracy of levels ($N_L = eB/h $).
As a result, some macroscopic parameters which depend on the total number of conduction electrons can only change in (the appropriately scaled) multiples of the degeneracy of the Landau levels. One particular effect seemingly related with this mechanism is known as the Quantum Hall Effect (QHE), which is quantization of the transversal conductance (ratio of the longitudinal current to the transversal voltage). Not that a full explanation of the QHE is constructed with arguments using this picture, but at least it is possible to come in touch with the QHE by using this type of reasoning. However, this mechanism alone becomes drastically insufficient as we attempt to explain the Fractional Quantum Hall Effect (FQHE). During the last two decades, researchers have proposed many new concepts and carried out a lot of calculations to explain the FQHE (e.g. cf. \cite{Chak}), including composite-particle type approach to the many-body Schr\"{o}dinger equation or work on the so-called localized states, or the so-called effective field theories. This latter development  (cf. \cite{Kivelson}) is based on the postulate that the 2D electron gas may interact with the magnetic field in some profoundly different ways than a single free electron does. In other words, in addition to the Lorentz force and Landau quantization, there may be another effect at play which is switched on in a planar electronic system under certain conditions. I subscribe to this idea, and will propose a mechanism for exactly that via the Mesoscopic Mechanics formulated in this article. It needs to be emphasized that the mechanism described by the MeM is utterly different than the effective field theory mentioned above. This ``new" mechanism of interaction does not by any means ``switch off" the Landau states, whenever these are permitted to form in a material, but rather it is an additional and separate effect that needs to be considered. The defining feature of the effect is that it gives rise to inhomogeneous distribution of the magnetic field throughout the sample. The exact form of this effect will be postulated and explained in Section \ref{MeM}.
 
As we will see, the dynamic variable of the Mesoscopic Mechanics introduced below is a transform. Before anything else, I would like to point out that this by itself is nothing very unusual for several reasons. First, it seems unavoidable in this kind of work to have some sort of an object that would account for global effects resulting from local interactions, e.g. some sort of an order parameter, and the transform postulated by the MeM fulfills this exact role.  Also worth mentioning here is the classical idea that the response of a macroscopic parameter to the external field could be viewed via the linear response model. This has been tried in the context of the Quantum Hall Effects via the well-known Kubo formula, which gives a treatment of the Hall current based on the Ansatz that it will respond linearly to the external field. In that approach conductivity is a multiplier---an object not unlike the operator which is the dynamic variable of the (quite nonlinear) Mesoscopic Mechanics. Finally, as we know scattering phenomena may be viewed as transforms.  Indeed, as the Hamiltonian is perturbed and the originally distinguished basis of eigenstates is replaced by another, the whole process may be encoded in the corresponding change-of-basis transform, even if typically such a transform would not be determined uniquely. A question stands, can this be understood from a higher level, i.e. is there a meta-theory that would take the transform itself as the dynamic variable and explain its particular value as the critical point of a meta-Hamiltonian appropriate for a given scattering process? Naturally, this is a question about mathematical structure of the physical theory rather than a problem of physics, which is not to say that it wouldn't be of interest from the purely physical standpoint. Anyhow, such a possibility is not unthinkable in general, and is interesting to mention in the context of the Mesoscopic Mechanics.

Finally, I concede it may yet turn out that the phenomenon I conjecture in this article may in fact be in some way derived from the many-body Schr\"{o}dinger picture. Either way, the MeM is of interest.

\section{Formulation of the Mesoscopic Mechanics}
\label{MeM}

The inspiration for the formulation of Mesoscopic Mechanics comes from a long work on the Nonlinear Maxwell Theory synoptically described in Section \ref{NM}. A prominent role in that latter theory is played by the deceivingly benign logarithmic integral $\int \ln f$ of a real function $f$. Remarkably, this expression has the meaning of the logarithm of the determinant of the operator of multiplication by $f$. Indeed, using the integral  seems to be the only correct way to re-normalize an otherwise divergent expression and make sense of the determinant of this operator. This algebraic object is in turn related to the entropy associated with an operator. Entropy has yet another description as the logarithm of the corresponding \emph{partition function}, say,
\begin{equation}
\log \int [\emph{D}\varphi ] \exp{(-\frac{1}{2}\langle K\varphi,K\varphi\rangle})
\label{entropy}
\end{equation}
Strictly speaking, this is the \emph{entropy} associated with the operator $KK^*$ rather than $K$ itself.  This ubiquitous in the Quantum Field Theory integral over the infinite-dimensional space is understood in any way suitable and will cause no essential technical difficulties in the context of our discussion.   

We now turn attention to an idealized planar electronic system exposed to the perpendicular magnetic field with magnetic induction $B$. We omit the appropriate constants, but adopt the convention that $B^2$ is measured in units of energy. Suppose the electronic system is characterized by the single-particle Hamiltonian $H$ whose exact nature is not specified \emph{a priori}. For example, $H$ could incorporate a periodic potential resulting in the Bloch states, or it could be the Landau Hamiltonian resulting in the Landau states, or it could incorporate impurity potentials possibly leading to localized states, etc. 

To steer the discussion away from mathematical technicalities, let us assume that the Hamiltonian $H$ has a discrete spectrum. Let $|\psi _n\rangle$ denote the complete set of states with the corresponding eigenvalues $E_n$, so that
\begin{equation}
H|\psi _n\rangle = E_n |\psi _n\rangle .
\label{Schrod}
\end{equation}
Here, the index $n$ is not a physical quantum number but a label indexing the eigenstates, with the convenient proviso that the corresponding energy $E_n$ is a nondecreasing function of $n$. 

Now, we introduce the \emph{complete} Hamiltonian $\Xi$ whose arguments are operators denoted $K$. It is defined as follows 
\begin{equation}
\Xi (K) = \mbox{trace} \left(K H K^* \right) + B^2\log \overline{\det}\left(KK^* \right)
\label{Ksi}
\end{equation}
Here $\overline{\det}$ denotes the determinant of the nondegenerate part, i.e. the product of all \emph{nonzero} eigenvalues accounting for their multiplicities. In fact, without loss of generality, as it turns out a posteriori, we may assume that 
\[
K:F\rightarrow G
\]
 is an operator with null kernel, $\ker K = \{0\},$ whose domain $F$ and target space $G = \mbox{Im}(K)$ are finite-dimensional subspaces of $\mbox{span}_{L_2}\left\{|\psi _n\rangle: \mbox{all }n\right\}$. Disregarding some constants, the second term on the right-hand side is regarded as essentially identical with the entropy (\ref{entropy}). It is intuitively appealing to say that the first term of the Hamiltonian $\Xi$ is responsible for a single-electron portion of the energy, while the \emph{entropy} term accounts for the energy of inter-electron interaction. The interaction is switched on with an application of the magnetic induction. It may be facilitated by fields of a predetermined character whose actual nature does not affect the theory in any way. Yet one may try and consider some more concrete scenarios, e.g. the fields $\varphi$ in (\ref{entropy}) could represent phonons, charge-waves, or spin-waves. As a matter of fact, this last possibility would require a spin-formulation of the theory which will be briefly addressed later. Finally, I do not exclude the possibility that the Hamiltonian $\Xi$ expresses an emergent \emph{fundamental} law, i.e. that this is how electron gas interacts with the ambient magnetic field even in the absence of any additional structures, e.g. even in the absence of the containing crystal lattice if it were at all feasible. As we will see, the entropy interpretation of the second part of $\Xi$ goes hand in hand with the probabilistic interpretation of $KK^*$ postulated below. Naturally, the model is more general than any single underlying physical system and there may be other applications and interpretations.

Let us consider extrema of the functional (\ref{Ksi}) subject to the constraint 
\[
\mbox{trace}\left(KK^*\right) = \mbox{const}. 
\]
A direct calculation shows that the critical points satisfy the Euler-Lagrange equation in the form
\begin{equation}
KH  +B^2 (K^*)^{-1} = \nu K.
\label{Euler-Lagrange}
\end{equation}
[The reader who carries out the calculation will see that there is also another equation, equivalent to this one via conjugation.] It follows that
\begin{equation}
K^*K=\frac{B^2}{\nu-H|_{H<\nu}},
\label{KstarK}
\end{equation}
where the restriction $|_{H<\nu}$ denotes the orthogonal projection to the subspace spanned by the eigenfunctions of $H$ corresponding to the eigenvalues strictly less than $\nu$. Therefore, any two solutions $K$ differ by a unitary transformation, say, $U$, and the general solution has the form
\begin{equation}
K=U\frac{B}{(\nu-H|_{H<\nu})^{1/2}}, 
\label{solutionK}
\end{equation}
or more explicitly
\begin{equation}
K = U \circ\sum_{E_n < \nu}\frac{B}{(\nu-E_n)^{1/2}}\quad |\psi _n\rangle\langle \psi_n|.
\label{theK}
\end{equation}
where 
\[U:F\rightarrow G,\qquad U^{-1}=U^*
\] 
is a unitary operator whose domain is 
\begin{equation}
F=\{H<\nu\}=\mbox{span}\{ |\psi _n\rangle : E_n < \nu \}. 
\label{theU}
\end{equation}
The space $F$ is interpreted as the Fermi sea at $T=0$ and remains fixed at all times. On the other hand, the target space $G$ is \emph{a priori} unspecified.
It has to be emphasized that
\[
K:F\rightarrow G
\]
is dimensionless.

It is perhaps worth pointing out that setting $B=0$ in the equation (\ref{Euler-Lagrange}) forces $\nu$ to become an eigenvalue and $K$ a generalized eigenstate $K=\Lambda_{\nu}$, i.e. an orthogonal projection onto the space spanned by all eigenstates corresponding to the eigenvalue $\nu$. It is remarkable that since the solution $K$ in (\ref{solutionK}) depends algebraically on the Hamiltonian $H$, all essential analytical difficulties are concealed in the treatment of the \emph{linear} operator $H$. This gives us total freedom in the choice of the type of problem we want to consider, e.g. a boundary value problem, etc.

With the basic notions already in place, it is clear that we have entered the domain of a new paradigm and there could be no interpretation of $K$ and $\Xi$ within the framework of a preexisting theory. This venture is here seen as necessary in order to understand some emergent phenomena encountered in planar electronic systems. To be sure, the Hamiltonian $\Xi$ has been concocted with the familiar elements of Quantum Mechanics and the Quantum Field Theory. It is only the relation of the operator $K$ to the physical system that needs to be postulated. With this understood, I will now put forward some ways of interpreting the $\Xi$ model. The emerging new paradigm is in harmony with the principles of Quantum Mechanics, accepting and building upon its interpretation and postulates. However, the interpretation of Mesoscopic Mechanics requires new postulates that are extrinsic to Quantum Mechanics.
\vspace{.5cm}

\noindent
\textbf{First Postulate of the MeM:}
{\em Suppose a magnetic field with magnetic induction $B\neq 0$ is applied transversally to a two-dimensional electron gas. Then, for an (``electronic") statistical state $W$ there is a corresponding (``magnetic") state $KWK^*$, where $K$ is a critical point of the Hamiltonian (\ref{Ksi}) given in (\ref{solutionK}) and (\ref{theK}). 
It is postulated that an observable $A$ representing a measurement of the magnetic field or its effects has the expectation
\begin{equation}
\langle A \rangle = \frac{\mbox{trace} (AKWK^*)}{\mbox{trace} (KWK^*)}.
\label{KWKexpectation}
\end{equation}
}

Observe that since the new state $KWK^*$ can be normalized, e.g. so as to guarantee 
\[
\mbox{trace} (KWK^*) =1,
\] 
it is in fact \emph{independent of the value of} $B$ as long as $B\neq 0$. Observe that only the below-Fermi part of the input state $W$ affects the output state $KWK^*$. 

We obtain an interesting example by applying this transform to the temperature-$T$ Fermi state 
\[
S=\int f(E,T)\Lambda_E\sqrt{E}dE, 
\] 
where 
\[
f(E,T) = 1/(\exp{\frac{E-\nu}{kT}}+1)
\]
is the Fermi distribution with Fermi energy $\nu$, and $\Lambda_E$ denotes the orthogonal projection on the space spanned by all the eigenstates of the Hamiltonian with eigenvalue $E$. The corresponding magnetic state (before normalization) will be
\[
KSK^* = U\circ \left(B^2\int\limits_0^{\nu}\frac{f(E,T)}{\nu-E}\Lambda_E\sqrt{E}dE\right)\circ U^*,
\]
Consider in particular the state $W=I|_{H<\nu}$ related to absolute zero temperature. In this case the magnetic state is 
\[
KK^* =  U\circ \sum_{E_n < \nu}\frac{B^2}{\nu-E_n}\quad |\psi _n\rangle\langle \psi_n|\circ U^*.
\]
Suppose for a moment that $U=I|_{H<\nu}$. Heuristically, we say that the respective probability of finding a magnetic flux quantum \emph{in residence} on the state $|\psi _n\rangle\langle \psi_n|$ is equal to 
\[
\frac{1}{\mbox{trace}(KK^*)}\quad \frac{B^2}{\nu-E_n}. 
\]
Consider another example in which $U$ consists in switching two states, say, $|\psi _k\rangle$ and $|\psi _l\rangle$ and acts as identity on the space spanned by the remaining states, i.e.
\[
U = \sum_{n \neq k,l}|\psi _n\rangle\langle \psi_n| +
|\psi _k\rangle\langle \psi_l| +
|\psi _l\rangle\langle \psi_k|.
\]
In this case
\[
KK^* = \sum_{n \neq k,l}\frac{B^2}{\nu-E_n}\quad |\psi _n\rangle\langle \psi_n| +
\frac{B^2}{\nu-E_k}\quad |\psi _l\rangle\langle \psi_l| +
\frac{B^2}{\nu-E_l}\quad |\psi _k\rangle\langle \psi_k|,
\]
i.e. we observe switching of the corresponding probabilities of finding the magnetic flux quanta in residence on states $|\psi _k\rangle\langle \psi_k|$ and $|\psi _l\rangle\langle \psi_l|$. This underscores the importance of understanding how $U$ may be allowed to evolve in time and depend on parameters, which will be discussed in Sections \ref{evol} and \ref{verif}. It seems tempting to think of 
\[
\frac{1}{\mbox{trace}(KK^*)}\quad \frac{B^2}{\nu-E_n}\quad |\psi _n\rangle\langle \psi_n|
\]
as a sort of elementary excitation as in, say, a Gedanken experiment in which the $n$'th electron evaporates above the Fermi level carrying away the corresponding fraction of the flux. One needs to keep in mind the probabilistic interpretation as well as the fact that this type of evolution can only be realized via the corresponding evolution of the transform $U$.

Let $K$ be a singular point of the Hamiltonian $\Xi$. It is interesting to observe that the  spectral characteristic of the magnetic state formed via the transform 
\begin{equation}
\label{transfrom}
W\longrightarrow KWK^*,
\end{equation}
is mainly determined by the electrons under the Fermi surface. The magnetic state is more than merely a virtual construction. In fact, very particular constraints are imposed on the possible outcomes of measurement of the magnetic field, e.g. an observable which is null on the subspace $G$ will return expected value zero in (\ref{KWKexpectation}). Separately, one should keep in mind that the occupation of the electronic states below the Fermi level remains unaffected. In particular, one should apply the electronic state in consideration of phenomena that are independent of the magnetic field, as for example the screening effects.

Since the First Postulate seems to fully determine how the magnetic properties of the system depend on its state, the Second Postulate which we will now formulate is perhaps not so much necessary as it is interesting. However, we note that so far the formulation of Mesoscopic Mechanics has been completely space-independent, i.e. fully contained within the framework of operator algebra while the electronic states played only an auxiliary role. Since the theory pertains to mesoscopic-scale phenomena after all, it would be incomplete without some indication of what is to be expected as regards the planar distribution of, say, the magnetic flux. The Second Postulate fulfills this specific function. It pertains to coherent states whose role in the many-particle setting is less pronounced and perhaps not so well understood as their single-particle applications. Anyhow, observe that to any electronic coherent state $C = \sum_n c_n |\psi_n\rangle$ we can assign a magnetic flux coherent state via
\begin{equation}
C\longrightarrow KC.
\label{coher}
\end{equation}
Again, both $C$ and $KC$ may require a specific normalization depending on an application. 
\vspace{.5cm}

\noindent
\textbf{Second Postulate of the MeM:}
{\em Consider the coherent state 
\[
\Psi = \sum_{\mbox{\small{filled states}}}  |\psi_n\rangle,
\] 
keeping in mind that in fact it depends on the phases of eigenstates. Suppose now that a magnetic field with magnetic induction $B\neq 0$ is applied transversally to a two-dimensional electron gas. It is postulated that the planar concentration of the magnetic flux is characterized by the coherent flux state 
\[K\Psi,
\]
where $K$ is a critical point of the Hamiltonian (\ref{Ksi}) given in (\ref{theK}). In particular, a measurement of the surface distribution of the magnetic flux is expected to be well approximated by 
\[
x\rightarrow \Phi|K\Psi |^2(x),
\] 
where $\Phi$ is the total magnetic flux through the surface.
}

Suppose for example that
\[K = B\sum_{E_n < \nu}\frac{\exp{i\varphi_n}}{(\nu-E_n)^{1/2}}\quad |\psi _n\rangle\langle \psi_n|,
\]
so that here the operator $U$ is diagonal in the basis of eigenstates. In such a case an application of the transform $K$ results in 
\begin{equation}
\label{fluxes}
K\Psi= B\sum_{\mbox{\small{filled states}}}\frac{\exp{i\varphi_n}}{(\nu-E_n)^{1/2}}\quad |\psi_n\rangle.
\end{equation}
If we agree that in this model $|\Psi |^2$ represents a planar concentration of electron charge (which would depend on the phases of the eigenstates !) then we may view $\Psi \rightarrow K\Psi$ as a charge-to-flux transform of sorts. The Figure shows two particular examples of the modulus function $|K\Psi |^2$ (in arbitrary scaling) in a free 2D electron gas at zero temperature. The effective wavelengths of the wave functions in a free electron gas model with periodic boundary conditions depend on the size of the sample (torus). In consequence we observe that in the \emph{correlated-phases mode}, i.e. when all $\varphi_n$'s are equal and all the phases of the states $|\psi_n\rangle$ are also equal, there is exactly one vortex on the whole torus. We observe a characteristic splitting and deformation of this vortex into a bunch of stripes when the phase modes $\varphi_n$ or the phases of the states $|\psi_n\rangle$ are distributed randomly. The fact not to be missed is that here the vortices are formed in total absence of the Landau states. In a real material, the spectral properties and the average separation between the vortices would depend explicitly and very heavily on the band structure below the Fermi surface, the particular waveforms of the corresponding electronic states, as well as the phase modes, including the $\varphi_n$'s. Also, unless the Fermi surface does not prohibit closed orbits thus suppressing formation of Landau states, as it may be the case in a strictly two-dimensional crystal for example, one should consider the Landau Hamiltonian.

Thus, the Hamiltonian (\ref{Ksi}) together with the transforms (\ref{transfrom}) and (\ref{coher}) provide a framework for the description of a new type of interaction of the magnetic field with the Fermi sea which is believed to universally occur in principally two-dimensional systems. The suggested interaction is independent and separate from the phenomenon of formation of Landau states.  Naturally, the observed effects of this phenomenon strongly depend on the band structure of the actual material and other parameters, like temperature but also the initial state $U_0$ and time evolution of the unitary transform $U$, which is discussed in Section \ref{evol}. In particular, a possibility of there being an energy gap at the Fermi level is of consequence for the nature of the transform $K$. Indeed, if there is no energy gap at all, then electrons occupying states strictly below the Fermi level will contribute only very weekly to the flux-density state. Informally speaking, the probability weights $B^2/(\nu-E_n)$ corresponding to the states from strictly below the Fermi level will be negligibly small as compared to the infinite weights falling on those states for which $E_n \simeq \nu$. If on the other hand, an energy gap $\triangle$ separates the Fermi level from the occupied states, then the corresponding distribution of weights will be more uniform. This sensitizes the theory to all the phenomena and material properties dependent on the existence and size of the energy gap, e.g. metal-insulator (Peierls) transition, energy gap in semiconductors, energy gap in superconductors, etc. Naturally, the presence of an energy gap makes the theory more sensitive to the entire band structure of a given material. It is a formidable yet worthwhile task to analyze the implications of this theory in more realistic band-structure models.

The emerging picture of MeM is that of a \emph{meta-theory}. The phenomena it describes result from the band-structure but do not affect it. Strictly speaking, it is only the sub-Fermi surface part of the band structure that is unaffected as the formation of the magnetic states should typically affect electrons with energies at the Fermi level. We do not attempt a detailed analysis of this latter problem here. A detailed analysis of this problem, e.g. description of the behavior of a free electron in the resulting nonuniform magnetic field would shed some light on how the phenomenon at hand may affect the Hall effect. The future may hold the solution of this fascinating problem.

\section{Time evolution}
\label{evol}

The presence of an analogy between two theories is always a nontrivial matter as it opens the possibility that the two theories may be just different facets of a yet unknown unifying higher-level construction. In this section we will postulate that the already clear analogy between the Mesoscopic Mechanics and the Schr\"{o}dinger Mechanics extends to time evolution as well. First, just as the matrix $U$ in (\ref{solutionK}) can a priori depend on a parameter, it is also free to depend on time. In view of the interpretation provided in the First Postulate of the MeM, if the system were conservative, the evolution of the state would be generated by the Hamiltonian. However, our matrix $U$ is finite dimensional and so cannot be obtained by exponentiating the infinite-dimensional Hamiltonian $H$ for one thing, and moreover the system as a whole is not characterized by $H$. Let us observe that when
\begin{equation}
U = U_0\exp{(i\nu t/\hbar)},
\label{Uevol}
\end{equation}
then the corresponding $K$ as in (\ref{solutionK}) is a solution of the Mesoscopic Schr\"{o}dinger equation
\begin{equation}
\label{MesoSchr}
i\hbar\dot{K} = -KH  -B^2 (K^*)^{-1} 
\end{equation}
One may interpret the solutions of this type as representing \emph{correlated evolution} in a certain sense as explained henceforth. Indeed, equation (\ref{MesoSchr}) admits other types of solutions, say, of the form 
\begin{equation}
K = \sum_{E_n < \nu}a_n(t)\quad |\psi _n\rangle\langle \psi_n|.
\label{async}
\end{equation}
All we need to guarantee is that all the $a_n$ satisfy the ordinary differential equation
\[
i\hbar\dot{a_n} = -E_n a_n  - \frac{B^2} {a_n^*}. 
\]
Writing $a_n=r_ne^{i\varphi_n}$, plugging it into the equation above and separating the real and imaginary parts we obtain
\[
\dot{r_n} = 0\quad\mbox{and}\quad \dot{\varphi_n} = (E_n + \frac{B^2} {r_n^2})/\hbar ,
\]
which implies
\begin{equation}
r_n = r_{n,0}\quad\mbox{and}\quad \varphi_n = \frac{1}{\hbar}\left(E_n + \frac{B^2} {r_{n,0}^2}\right)t +\varphi_{n,0}.
\label{a_n}
\end{equation}
In consequence, the resulting $K$ as in (\ref{async}) will typically not correspond to the critical points of the functional $\Xi$ given in (\ref{Ksi}). However, in such a case the $a_n$'s will oscillate each with a different frequency, which we may interpret as an uncorrelated evolution of the corresponding components (or electrons). When all $a_n$'s are in sync at all times, and oscillate with the frequency, say, $\nu$, then (\ref{a_n}) shows that necessarily
\[
r_n = \frac{\pm B}{(\nu - E_n)^{1/2}}
\]
and so we are again in the regime (\ref{theK}), and so $K$ is a critical point of the functional $\Xi $. This is a complementary phenomenon to that of phase correlation, which has been discussed in the previous section and illustrated in the figures. In particular, the phase-correlated regime will remain such when the evolution follows the pattern prescribed in (\ref{MesoSchr}).

As a digression, it is interesting to note in the context of (\ref{Uevol}) that at least in the free electron model, when the system is in a state $KSK^*$ then the expected value of the single-particle Hamiltonian is 
\[
\langle H \rangle = \mbox{trace} (HKSK^*)=\nu -\delta(\triangle ),
\] 
where the $\delta(\triangle )$ depends on the energy gap $\triangle$ at the Fermi level, although it is not equal to it, and $\delta\longrightarrow 0$ when  $\triangle\longrightarrow 0$. This is verified by a direct calculation via the continuous approximation in the momentum space. 

It is also interesting to observe that in principle the $\hbar$ in (\ref{MesoSchr}) could represent a Hermitian matrix (of the same dimension as $K$). However, I do not see any application for this latter fact at present.

\section{Constraints, extensions and verifiability}
\label{verif}

A question arises as to whether there may be external physical constraints on the unitary part $U$ of the transform $K$. The Figure demonstrates that two solutions corresponding to the same single-particle Hamiltonian, yet a different selection of the unitary component, say $U\in U(N)$, will have significantly different physical properties. One would like to know how such different solutions can be realized in a physical system. It seems natural to expect that different states may be prepared via a cyclic perturbation of the single-particle Hamiltonian. What I have in mind here is quite similar in spirit to the phenomenon of the Berry phase (cf. \cite{Bohm}) and what we have learned from it. In other words, one needs to consider a parameter space indexing the single-particle Hamiltonian. As one walks along a loop in the parameter space, the evolution equation (\ref{MesoSchr}) forces the corresponding states $K$ to trace a path in the total space of a $U(N)$-principal bundle. In analogy to the Berry phase theory, the (non-Abelian !) bundle is endowed with a natural geometry, i.e. a principal connection and its curvature, which are determined by the fine properties of the perturbation of the Hamiltonian and the resulting evolution of states. A careful look at the various holonomy questions in this geometry may bring answers as to the constraints on the possible values of the non-Abelian phase and its stability. Depending on the answers, this point may have a variety of interesting implications and applications in materials engineering. A study of feasible perturbations and a construction of a suitable geometric formalism to describe such non-Abelian phase phenomena in the context of Mesoscopic Mechanics will be attempted in the future, circumstances permitting.

As regards the problem of verifying the MeM experimentally several routes could be taken, even now. As we have pointed out already, the picture of the magnetic vortex obtained in the MeM depends explicitly on the characteristic of the material, which  fact opens plethora of natural questions for experiment as well as theory. The magnetic vortices arising in the MeM have a very definite spectral profile, which could, at least in principle, be verified experimentally. Moreover, due to the particular form of the operator $K$, the MeM can be attuned to perturbation analysis, e.g. via the Lippman-Schwinger type approach. Now, it seems quite realistic to try and compare  predictions of the MeM with experiment for carefully designed scattering and other perturbation experiments. The predictions we have in mind pertain to the spectral profile of the magnetic flux, but also its effects, e.g. on the electrons at the Fermi level.

Finally, I would like to briefly signal that the entire framework of the Mesoscopic Mechanics admits a natural generalization to the noncommutative setting that would incorporate the electron spin into the picture. To let on the crux of the matter, replacing the phase factors in formula (\ref{theK}) by a collection of unitary, say, $2$-by-$2$ matrices still yields a solution of the Euler-Lagrange equation (\ref{Euler-Lagrange}). This leads to some immensely interesting questions. Separately, just as the original formulation of the MeM presented here parallels the Schr\"{o}dinger mechanics, the theory also admits a relativistic formulation in parallel to the Klein-Gordon type setting. It is important to ask if there is also a Dirac type relativistic formulation. I have no complete answers to all these questions at present.

\section{The role of the NMT}
\label{NM}

There are a few reasons to evoke here some highlights of the (fully) Nonlinear Maxwell Theory. First, as already explained the NMT leads to the Mesoscopic Mechanics, and in a way the latter is deduced from the former. Secondly, it is worthwhile to realize that the MeM is part of a broader framework that has already been shown to provide a phenomenological model of some landmark low-temperature phenomena:
\begin{itemize}
\item
magnetic vortex lattice (cf. \cite{sowa3})
\item
magnetic oscillations (cf. \cite{sowa1})
\item
charge stripes (cf. \cite{sowa4})
\item
superconductivity (cf. \cite{sowa4})
\end{itemize}
Yet another reason is to announce that the NMT has been tied to the fundamental principles via the Mesoscopic Mechanics. 

The Nonlinear Maxwell Equations couple the electric and the magnetic fields (resp. $\vec{E}$ and $\vec{B}$) to a scalar real-valued field variable $f$. In a certain sense $f$ is dual to the dynamic variable $K$ of the Mesoscopic Mechanics. 
When we are allowed to assume that the electric and magnetic field vectors are not perpendicular 
\begin{equation}
\vec{E}\cdot\vec{B} \neq 0, 
\label{correlation}
\end{equation}
the Nonlinear Maxwell Equations can be rewritten in an especially interesting form:
\begin{equation}
\frac{\partial\vec{B}}{\partial t}+\nabla\times\vec{E}=0
\label{first}
\end{equation}
\begin{equation}
\nabla\cdot\vec{B} =0
\label{second}
\end{equation}
\begin{equation}
(\frac{\partial\vec{E}}{\partial t}-\nabla\times\vec{B})\times\vec{E}
+(\nabla\cdot\vec{E})\vec{B} = -(\vec{E}\cdot\vec{B})\nabla\ln f
\label{third}
\end{equation}
\begin{equation}
\label{fourth}
(\frac{\partial\vec{E}}{\partial t}-\nabla\times\vec{B})\cdot\vec{B}
= -(\vec{E}\cdot\vec{B})\frac{\partial}{\partial t}\ln f
\end{equation}
\begin{equation}
(\frac{\partial ^2}{\partial t^2} - \bigtriangleup )
f + (|\vec{B}|^2 - |\vec{E}|^2)f = \nu f.
\label{last}
\end{equation}
Note that the term $(\nabla\cdot\vec{E})\vec{B}$ featured in the equation (\ref{third}) has the meaning of the magnetic flux modulated by (``residing on") the electric charge. This concept pervades the whole theory, including the Mesoscopic Mechanics. It is easily seen that in two spacial dimensions the system can be reduced to a single nonlinear scalar equation:
\begin{equation}
-\bigtriangleup f(x,y) +\frac{B^2}{f(x,y)} =\nu f(x,y),
\label{scal2d}
\end{equation}
which is the Euler-Lagrange equation for the critical points of the functional  \begin{equation}
L(f)= \frac{1}{2}\int |\nabla f|^2 + B^2\int\ln(f)
\label{functional}
\end{equation}
subject to the constraint:
\begin{equation}
\int f^2  = \mbox{const}.
\label{constraint}
\end{equation}
The functional $L$ is neither bounded below nor above, so that one is looking at
the problem of existence of {\em local} extrema. This functional was studied via a custom-designed asymptotically stable discrete approach in \cite{sowa3}. The functional (\ref{functional}) is the precursor of (\ref{Ksi}), where the function $f$ is replaced with an operator $K$. The description of the magnetic vortex lattice given in \cite{sowa3} is a classical counterpart of what has been presented in Section \ref{MeM}.

The spotting of one of the basic features of magnetic oscillations within the framework of the NMT is a beautiful and somewhat mysterious phenomenon worthy a longer comment.  As already mentioned in Section \ref{emerge}, condensation of electrons at the Landau levels in conjunction with the Fermi surface crossing result in magnetic oscillations, e.g. the longitudinal resistivity in the QHE experiment undergoes quantum oscillations that are in a certain way correlated with the plateaus of Hall resistance.  To obtain a quantitative picture of the oscillations (\emph{in metals}) one needs to calculate the thermodynamic potential and observe its dependence on the energy levels as the magnetic field is switched on. This results in the so-called  Lifshitz-Kosevich formula (cf. \cite{Sch}). Experiment shows that the oscillating macroscopic parameter invariably displays a characteristic distorted-sinusoidal pattern. However, this fact cannot be accounted for by the Landau-Fermi picture itself, and neither does it follow from the said formula, but is justified via the ferromagnetic feedback and so it requires the assumption of ferromagnetism. Quite surprisingly, the same pattern occurring in much the same context is intrinsically present in the Nonlinear Maxwell Theory \cite{sowa1}, which of course has nothing to do with the ad hoc argument from ferromagnetism. 

Finally, it needs to be emphasized that the Mesoscopic Mechanics \emph{is not} the result of canonical quantization of the Nonlinear Maxwell Theory. I do not consider the latter task in this article but I think a brief sketch of what would be involved is appropriate for the sake of completeness. The system of equations (\ref{first})-(\ref{last}) is obtained from gauge-theoretic equations:
\begin{equation}
 dF_{A}=0
\label{syst0}
\end{equation}
\begin{equation}
 \delta (fF_{A})=0
\label{syst1}
\end{equation}
\begin{equation}
 \Box f +|F_{A}|^{2}f=\nu f.
\label{syst2}
\end{equation}
where $A$ is the electromagnetic vector potential, so that the corresponding electromagnetic field is $ F_A = dA$. In particular, when the system is written in this form the assumption (\ref{correlation}) is no longer required. It is important to realize that the equations (\ref{syst0})-(\ref{syst2}) are not of the Euler-Lagrange type for any Lagrangian (cf. \cite{sowa4}). This may at a first glance appear un-physical, and so deserves a more detailed comment. An interesting idea one could try and pursue is that the equations may be completed to an Euler-Lagrange system by coupling them to an additional field, but this is not a solution I would like to put forward here. A more direct possibility is in that the equations can be deformed in a continuous (\emph{adiabatic}) manner so that the resulting system will in fact correspond to the critical points of a certain Lagrangian, and additionally the deformed system will have the same two-dimensional reduction (\ref{scal2d}). These objectives are all achieved by the following simple trick. Use an auxiliary function 
\[
\varphi (x) = (c-\ln{|x|})^{-1}
\]
for a constant $c$, and consider the functional
\[
\int \varphi (f)|F_A|^2+\int|\nabla_{(t,x)}f|^2
\] 
subject to the constraint $\int f^2 = const$. A direct calculation shows that the critical points of this functional satisfy a system of equations similar to (\ref{syst0})-(\ref{syst2}). In fact, I predict that this new system of equations is indeed an \emph{adiabatic deformation} of the original equations, and its solutions display closely similar behavior to (\ref{syst0})-(\ref{syst2}). The point is that the function $f$ assumes values in a bounded interval in physically interesting solutions of the NM and so, for a suitable choice of the constants, $f$ will be well approximated by $\varphi (f)$. Moreover, a direct calculation shows that the deformed system has the same two-dimensional reduction (\ref{scal2d}). That the adiabatically deformed system can be quantized in a canonical way is a fact of significance.

\newpage
\noindent
\textbf{Fig.} Luminance graphs of the modulus function $|K\Psi |^2$ for $K\Psi$ prescribed in (\ref{fluxes}) with correlated phases (top) and uncorrelated phases (bottom).

\vbox spread 1.5in{}
\includegraphics{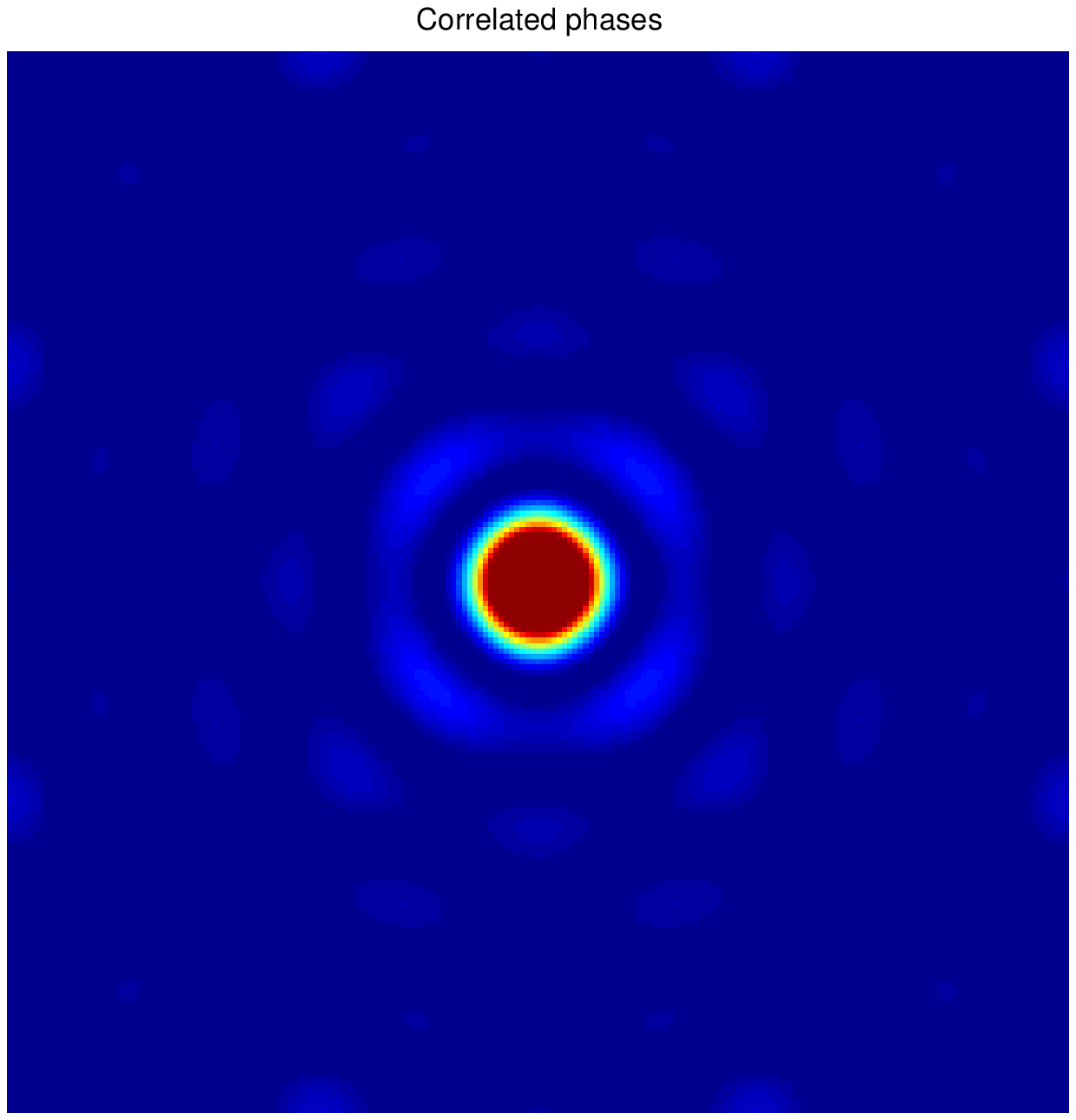}
\vbox spread 1.5in{}
\includegraphics{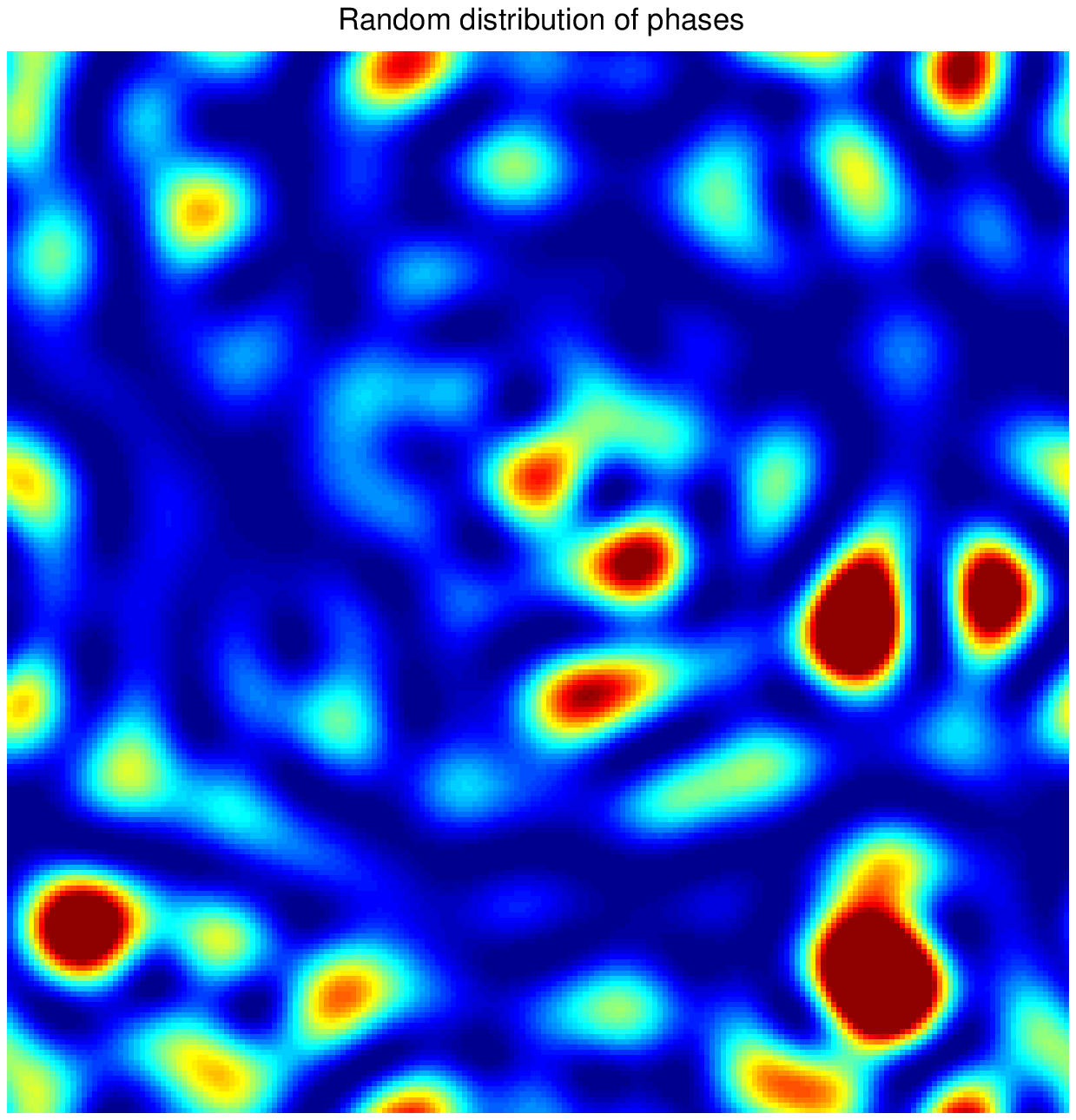}

\end{document}